\documentclass[12pt]{article}
\usepackage[dvips]{color}
\usepackage{epsfig}
\usepackage{amsmath}
\usepackage{txfonts}
\usepackage{graphicx}
\def\Box{\hbox{$\rlap{$\sqcup$}\sqcap$}}
\begin{document}
\title{\bf The Ultraviolet and Infrared Behavior of an Abelian Proca Model From the Viewpoint of a One-Parameter Extension of the Covariant Heisenberg Algebra }

\author{M. Ranaiy, S. K. Moayedi \thanks{Corresponding author, E-mail:
s-moayedi@araku.ac.ir}\hspace{1mm}\\
{\small {\em  Department of Physics, Faculty of Sciences,
Arak University, Arak 38156-8-8349, Iran}}\\
}

\date{\small{}}
\maketitle
\begin{abstract}
\noindent Recently a one-parameter extension of the covariant
Heisenberg algebra with the extension parameter $l$ ($l$ is a
non-negative constant parameter which has a dimension of
$[momentum]^{-1}$) in a $(D+1)$-dimensional Minkowski space-time has
been presented [G. P. de Brito, P. I. C. Caneda, Y. M. P. Gomes, J.
T. Guaitolini Junior and V. Nikoofard, Effective models of quantum
gravity induced by Planck scale modifications in the covariant
quantum algebra, \textit{Adv. High Energy Phys.} \textbf{2017}
(2017) 4768341]. The Abelian Proca model is reformulated from the viewpoint of the
above one-parameter extension of the covariant Heisenberg algebra. It is shown that the free space solutions of the above
modified Proca model satisfy the modified dispersion relation
$\frac{\textbf{p}^2}{\left(1+\frac{\Lambda^2}{2\hbar^2}\textbf{p}^2\right)^2}=m^2c^2$
where $\Lambda=\hbar l$ is the characteristic length scale in our
model. This modified dispersion relation describes two massive
vector particles with the effective masses ${\cal
M}_{_\pm}(\Lambda)=\frac{2m}{1\mp\sqrt{1-2\left(\frac{mc\Lambda}{\hbar}\right)^2}}$.
Numerical estimations show that the maximum value of $\Lambda$ in a
four-dimensional space-time is near to the electroweak length scale,
i.e., $\Lambda_{_{max}}\sim l_{_{electroweak}}\sim10^{-18}\; m$. We
show that in the infrared/large-distance domain the modified Proca
model behaves like an Abelian massive Lee-Wick model which has been
presented by Accioly and his co-workers [A. Accioly, J.
Helayel-Neto, G. Correia, G. Brito, J. de
 Almeida and W. Herdy, Interparticle potential energy for
D-dimensional electromagnetic models from the corresponding scalar
ones, \textit{Phys. Rev. D} \textbf{93} (2016) 105042].

\noindent
\hspace{0.35cm}



\end{abstract}

\section{Introduction}
Although quantum field theory is a very successful theory which
describes the fundamental interactions at the microscopic level, the
study of short-distance (high-energy) behavior of fundamental
interactions in quantum field theory leads to ultraviolet
divergences [1-3]. Today we know that these ultraviolet divergences
in quantum field theories can be removed by using the standard
renormalization techniques. One of these renormalization techniques
which is very close to the Pauli-Villars regularization technique is
the addition of higher-order derivative terms to the Lagrangian
density of a quantum field theory [1-3]. On the other hand, this
idea that there is a minimal length scale in the measurement of
space-time distances of the order of the Planck length is predicted
by different theories of quantum gravity such as string theory, loop
quantum gravity and non-commutative geometry [4-6]. The existence of
this minimal length scale in quantum gravity leads to the following
generalized uncertainty principle:
\begin{equation}
\Delta X\sim\frac{\hbar}{\Delta P}+\alpha'\frac{\Delta P}{\hbar},
\end{equation}
where $l_{_S}=\sqrt{\alpha'}\approx10^{-32}\; cm$, $l_{_S}$ is the
string length, and $\frac{\hbar c}{\alpha'}$ is the string tension
[4]. The generalized uncertainty principle (1) implies the existence
of a nonzero minimal length scale which is given by
\begin{equation}
\nonumber \Delta X_{_{min}}=2l_{_S}.
\end{equation}
It should be noted that the reformulation of the quantum field
theory in the presence of a minimal length scale is another way for
obtaining a divergence free quantum field theory [4-6]. In 2006, C.
Quesne and V. M. Tkachuk introduced a $(\beta,\beta')$-two-parameter
extension of the covariant Heisenberg algebra in a
$(D+1)$-dimensional Minkowski space-time which is described by the
following modified commutation relations:
\begin{eqnarray}
\nonumber \left[X^{\mu},P^{\nu}\right] &=&-
i\hbar\left[(1-\beta\textbf{P}^2)\eta^{\mu\nu}-\beta'P^{\mu}P^{\nu}\right],\\
\nonumber\left[X^{\mu},X^{\nu}\right]
&=&i\hbar\;\frac{2\beta-\beta'-(2\beta+\beta')\beta\textbf{P}^2}{1-\beta\textbf{P}^2}
\;(P^{\mu}X^{\nu}-P^{\nu}X^{\mu}),\\
\left[P^{\mu},P^{\nu}\right] &=& 0,
\end{eqnarray}
where $\mu,\nu=0,1,2,\cdots,D$, $\beta$ and $\beta'$
 are two non-negative constant parameters with dimension of
 $[momentum]^{-2}$, $X^{\mu}$ and $P^{\mu}$ are the modified position
 and momentum operators,
 $\textbf{P}^2=(P^0)^2-\sum^{D}_{i=1}(P^i)^2=(P^0)^2-\vec{\textbf{P}}^2$, and
 $\eta_{\mu\nu}=\eta^{\mu\nu}=diag(+1,\underbrace{-1,-1,\ldots,-1}_{D\;\;times})$
 is the flat Minkowski metric [6]. The Quesne-Tkachuk algebra (2)
 predicts the existence of an isotropic minimal length scale which
 is given by
\begin{equation}
\nonumber(\Delta X^i)_{_0}=\hbar
\sqrt{(D\beta+\beta')\left[1-\beta\langle(P^0)^2\rangle\right]},\qquad\forall
i\in\{1,2,\ldots,D\}.
\end{equation}
In recent years, reformulation of Maxwell electrodynamics in the
presence of a minimal length scale and the study of short-distance
behavior of Maxwell theory has attracted a considerable attention
among researchers in quantum field theory [7-10]. In Ref. [9], it
has been shown that in minimal length electrostatics the classical
self-energy of a point charge has a finite value. A free massless
spin-2 field $h_{\mu\nu}(x)$ in a $(D+1)$-dimensional Minkowski
space-time is described by the Pauli-Fierz action as follows:
\begin{subequations}
\begin{eqnarray}
S_{_{PF}}&=&\frac{c^3}{16\pi G_N^{(D+1)}}\int d^{D+1}x\;{\cal
L}_{_{PF}},\\
{\cal L}_{_{PF}}&=&\frac{1}{4}\left(\partial_\lambda
h^{\mu\nu}(x)\partial^\lambda h_{\mu\nu}(x)-2\partial_\mu
h^{\mu\nu}(x)\partial^\lambda h_{\lambda\nu}(x)+2\partial_\mu
h^{\mu\nu}(x)\partial_\nu h_{\;\lambda}^\lambda (x)-\partial_\mu
h_{\;\nu}^\nu (x)\partial^\mu h_{\;\lambda}^\lambda (x)\right).
\end{eqnarray}
\end{subequations}
The reformulation of the Pauli-Fierz theory from the viewpoint of
the Quesne-Tkachuk algebra has been studied in details in Ref. [11].
In Ref. [12], the following non-local model for electrodynamics has
been presented
\begin{equation}
{\cal L}=-\frac{1}{4\mu_0}\left(\;e^{l_*^{\;2}
\Box}\;F_{\mu\nu}(x)\right)\;\left(e^{l_*^{\;2}
\Box}\;F^{\mu\nu}(x)\right)-J^\mu(x)A_\mu(x),
\end{equation}
where $l_*$ is a constant parameter which has a dimension of
$[length]$, $A^\mu=(\frac{\phi}{c},A_x,A_y,A_z)$ is the potential
four-vector, $J^\mu=(c\rho,J_x,J_y,J_z)$ is the current four-vector,
and $\Box=\frac{1}{c^2}\frac{\partial^2}{\partial t^2}-\nabla^2$ is
the d'Alembertian operator in a four-dimensional flat space-time.
The authors of Ref. [12] have shown that the classical self-energy
of a point charge in Eq. (4) has a finite value (see Eq. (44) in
Ref. [12]). In a recent paper, a one-parameter extension of the
covariant Heisenberg algebra in a $(D+1)$-dimensional Minkowski
space-time has been suggested [13]. This algebra is a covariant
generalization of the Kempf-Mangano algebra [14]. In the present
paper the ultraviolet/short-distance and infrared/large-distance
behavior of an Abelian Proca model in the framework of the covariant
Kempf-Mangano algebra are studied analytically.

This paper is organized as follows. In Sect. 2, the structure of
one-parameter extension of the covariant Heisenberg algebra in a
$(D+1)$-dimensional flat space-time is introduced according to Ref.
[13]. In Sect. 3, Lagrangian reformulation of the Abelian Proca
model from the viewpoint of one-parameter extension of the covariant
Heisenberg algebra in a $(D+1)$-dimensional space-time is presented.
In Sect. 4, we show that the free space solutions of the modified
Proca model in Sect. 3 describe two massive vector particles. Our
calculations show that there is a characteristic length scale
$\Lambda$ whose maximum value is near to the electroweak length
scale, i.e., $\Lambda_{_{max}}\sim l_{_{electroweak}}\sim10^{-18}\;
m$. In summary and conclusions, we show that in the infrared region
the modified Proca theory in Sect. 3 behaves like an Abelian massive
Lee-Wick model. SI units are used throughout this paper.

\section{One-Parameter Extension of the Covariant Heisenberg Algebra}

In 2017, G. P. de Brito and co-workers introduced a one-parameter
extension of the covariant Heisenberg algebra [13]. This algebra in
a $(D+1)$-dimensional Minkowski space-time is characterized by the
following modified commutation relations:\footnote{It must be
emphasized that this algebra is a relativistic generalization of the
following algebra
\begin{eqnarray*}
\left[X^{i},P^{j}\right] &=&
i\hbar\left(\frac{l^2\vec{\textbf{P}}^2}{\sqrt{1+2l^2\vec{\textbf{P}}^2}-1}\delta^{ij}+l^2P^{i}P^{j}\right)\\
&=& i\hbar
\left(\frac{1+\sqrt{1+2l^2\vec{\textbf{P}}^2}}{2}\delta^{ij}+l^2P^{i}P^{j}\right),\\
\left[X^{i},X^{j}\right] &=& 0,\\
\left[P^{i},P^{j}\right] &=& 0,\qquad i,j=1,2,\cdots,D,
\end{eqnarray*}
which was introduced previously by Kempf and Mangano in Ref. [14].}

\begin{eqnarray}
\left[X^{\mu},P^{\nu}\right] &=&- i\hbar
\left(\frac{1+\sqrt{1-2l^2\textbf{P}^2}}{2}\eta^{\mu\nu}-l^2P^{\mu}P^{\nu}\right),\\
\left[X^{\mu},X^{\nu}\right] &=& 0,\\
\left[P^{\mu},P^{\nu}\right] &=& 0,
\end{eqnarray}
where $l$ is a non-negative constant parameter which has a dimension
of $\left[momentum\right]^{-1}$. The modified position and momentum
operators $X^{\mu}$ and $P^{\mu}$ in the above algebra have the
following exact coordinate representation [13]
\begin{eqnarray}
X^{\mu}&=&x^{\mu},\\
P^{\mu}&=&\frac{1}{1+\frac{l^2}{2}\mathbf{p}^2}p^{\mu},
\end{eqnarray}
where $x^{\mu}$ and $p^{\mu}=i\hbar\partial^{\mu}$ are the position and
momentum operators which satisfy the following usual covariant
Heisenberg algebra

\begin{eqnarray}
\left[x^{\mu},p^{\nu}\right] &=&- i\hbar \eta^{\mu\nu}, \\
\left[x^{\mu},x^{\nu}\right] &=& 0, \\
\left[p^{\mu},p^{\nu}\right] &=& 0.
\end{eqnarray}
In Eq. (9) $\textbf{p}^2=p_{\alpha} p^{\alpha}=-\hbar^2\Box$.\\
According to Eqs. (8) and (9) in order to reformulate a quantum
field theoretical model in the framework of a one-parameter
extension of the covariant Heisenberg algebra, the usual position and
derivative operators $(x^{\mu},\partial_{\mu})$ must be replaced by
the modified position and derivative operators
$(X^{\mu},\nabla_{\mu})$ as follows:
\begin{eqnarray}
x^\mu \longrightarrow  X^{\mu} &=&x^{\mu} , \\
\partial_\mu \longrightarrow
\nabla_{\mu}&:=&\frac{1}{1-\frac{(\hbar l)^2}{2}\Box}\partial_{\mu}.
\end{eqnarray}
Note that the quantity $\hbar l$ in Eq. (14) defines a
characteristic length scale $\Lambda:=\hbar l$ in our calculations.
For the space-time distances very greater than $\Lambda$
the space-time algebra becomes the usual covariant Heisenberg
algebra (Eqs. (10)-(12)), while for the space-time distances very smaller than $\Lambda$ the structure of
space-time must be described by Eqs. (5)-(7).

\section{Reformulation of the Abelian Proca Model in the Framework of One-Parameter Extension
of the Covariant Heisenberg Algebra}

The Abelian Proca model for a massive spin $1$ vector field $A^\mu
=(\frac{\phi }{c},A^1,\cdots,A^D)=(\frac{\phi }{c},\textbf{A})$ in
the presence of an external current $J^\mu=(c\rho
,J^1,\cdots,J^D)=(c\rho ,\textbf{J})$ in a $(D+1)$-dimensional
Minkowski space-time is [15-17]
\begin{equation}
{\cal L}=-\;\frac{1}{4\mu_{0}}F_{\mu \nu}(x)F^{\mu \nu}(x)+\frac{1}{2\mu
_0}\left(\frac{mc}{\hbar}\right)^2A_\mu(x) A^\mu(x)-J^\mu(x) A_\mu(x) ,
\end{equation}
where $F_{\mu \nu}=\partial_{\mu}A_\nu-\partial_{\nu}A_\mu$ is the
electromagnetic field tensor and $m$ is the mass of the gauge
particle. If we use Eq. (13) together with the transformation rule
for a covariant vector, we will obtain the following results
\begin{eqnarray}
\nonumber A_\mu (x) \longrightarrow  B_{\mu}(X) &=&\frac{\partial
x^\nu}{\partial X^{\mu}}A_\nu (x)\\ \nonumber  &=&\frac{\partial
x^\nu}{\partial x^{\mu}}A_\nu (x)  \\ \nonumber &=&\delta ^\nu _\mu
A_\nu (x)\\&=&A_\mu (x),
\end{eqnarray}
\begin{eqnarray}
\nonumber J_\mu (x) \longrightarrow  j_{\mu}(X) &=&\frac{\partial
x^\nu}{\partial X^{\mu}}J_\nu (x)\\ \nonumber  &=&\frac{\partial
x^\nu}{\partial x^{\mu}}J_\nu (x)  \\ \nonumber &=&\delta ^\nu _\mu
J_\nu (x)\\&=&J_\mu (x).
\end{eqnarray}
Using Eqs. (14) and (16) the modified electromagnetic field tensor
$G_{\mu \nu}(X)$ becomes
\begin{eqnarray}
\nonumber F_{\mu \nu} (x) \longrightarrow  G_{\mu \nu}(X)
&=&\nabla_{\mu}B_\nu(X)-\nabla_{\nu}B_\mu(X)\\
&=&\frac{1}{1-\frac{\Lambda^2}{2}\Box}F_{\mu \nu} (x).
\end{eqnarray}
If we use Eqs. (15)-(18), we will obtain the modified Lagrangian
density for an Abelian Proca model in the ultraviolet region as follows:
\begin{eqnarray}
\nonumber {\cal L}&=&-\;\frac{1}{4\mu_{0}}G_{\mu \nu}(X)G^{\mu
\nu}(X)+\frac{1}{2\mu _0}\left(\frac{mc}{\hbar}\right)^2B_\mu(X)
B^\mu(X)-j^\mu(X) B_\mu(X)\\ \nonumber &=&-\;\frac{1}{4\mu_{0}}\left(\frac{1}{1-\frac{\Lambda^2}{2}\Box}F_{\mu \nu} (x)\right)\left(\frac{1}{1-\frac{\Lambda^2}{2}\Box}F^{\mu \nu} (x)\right)+\frac{1}{2\mu _0}\left(\frac{mc}{\hbar}\right)^2A_\mu(x)
A^\mu(x)-J^\mu(x) A_\mu(x)
\\&=&\sum^{\infty}_{n=0}\Lambda^{2n}{\cal
L}_n,
\end{eqnarray}
where
\begin{eqnarray}
{\cal L}_0 &=&-\;\frac{1}{4\mu_{0}}F_{\mu \nu}(x)F^{\mu
\nu}(x)+\frac{1}{2\mu_0}\left(\frac{mc}{\hbar}\right)^2A_\mu(x)
A^\mu(x)- J^{\mu}(x) A_\mu(x),\\
{\cal L}_1&=&-\;\frac{1}{4\mu_ {0}} F_{\mu \nu}(x)\Box F^{\mu
\nu}(x).
\end{eqnarray}
The Lagrangian density (19) describes an infinite derivative Abelian
massive gauge field $A_\mu(x)$. It should be noted that the
expression ${\cal L}_0+\Lambda^2{\cal L}_1$ in the above equations
is the Lagrangian density of an Abelian
 massive Lee-Wick model
[18].\footnote{Lee-Wick model is
 a generalization of the usual quantum
electrodynamics in which mass renormalization, charge
renormalization, and wave function renormalization are finite
quantities [19,20].} For a classical field theory which is described
by the following Lagrangian density:
\begin{equation}
{\cal L}={\cal
L}(A_\lambda,\partial_{\nu_1}A_\lambda,\partial_{\nu_1}\partial_{\nu_2}A_\lambda,
\partial_{\nu_1}\partial_{\nu_2}\partial_{\nu_3}A_\lambda,\ldots),
\end{equation}
the Euler-Lagrange equation for the gauge field $A_\lambda$ becomes
[21,22]
\begin{equation}
\frac{\partial {\cal L}}{\partial A_\lambda}-\left(\frac{\partial
{\cal L}}{\partial
A_{\lambda,\nu_1}}\right)_{\nu_1}+\left(\frac{\partial {\cal
L}}{\partial A_{\lambda,\nu_1 \nu_2}}\right)_{\nu_1 \nu_2}-\ldots
+(-1)^k \left(\frac{\partial {\cal L}}{\partial A_{\lambda,\nu_1
\nu_2\ldots \nu_k}}\right)_{\nu_1 \nu_2 \ldots \nu_k}+\ldots =0,
\end{equation}
where
\begin{eqnarray}
 A_{\lambda,\nu_1\nu_2\ldots
 \nu_k}&:=&\partial_{\nu_1}\partial_{\nu_2}\ldots
 \partial_{\nu_k}A_\lambda,\\
\frac {\partial A_{\lambda,\nu_1\nu_2\ldots \nu_k}}{\partial
A_{\lambda,\mu_1\mu_2\ldots \mu_k}}&=&\delta ^{\mu_1}_{\nu_1}\delta
^{\mu_2}_{\nu_2}\ldots\delta ^{\mu_k}_{\nu_k}.
\end{eqnarray}
If we insert Eq. (19) into Eq. (23), we will obtain the
inhomogeneous infinite derivative Proca equation as follows:
\begin{equation}
\frac {1}{\left(1-\frac{\Lambda^2}{2}\Box\right)^2}\partial_\mu
F^{\mu\nu}(x)+\left(\frac{mc}{\hbar}\right)^2A^\nu(x)=\mu_0J^\nu(x).
\end{equation}
In the limit $\Lambda\rightarrow 0$, the modified Proca equation in
Eq. (26) becomes the usual Proca equation, i.e.,
\begin{equation}
\partial_\mu F^{\mu\nu}(x)+\left(\frac{mc}{\hbar}\right)^2A^\nu(x)=\mu_0J^\nu(x).
\end{equation}
After taking divergence of both sides of Eq. (26) and using
the relations $\partial_\nu\partial_\mu F^{\mu\nu}(x)=0$ and
$[\partial_\nu ,\Box]=0$, we obtain
\begin{equation}
\partial_\mu J^\mu(x)=\frac {1}{\mu_0}\left(\frac{mc}{\hbar}\right)^2
\partial_\mu A^\mu(x).
\end{equation}
Note that the above equation is a consequence of the modified field
equation (26). If we substitute (28) in (26), we will obtain
\begin{equation}
\frac {1}{\left(1-\frac{\Lambda^2}{2}\Box\right)^2}\Box A^\nu (x)+
\left(\frac{mc}{\hbar}\right)^2A^\nu(x)=\\\mu_0\left[J^\nu(x)+\left(\frac{\hbar}{mc}
\right)^2\frac
{1}{\left(1-\frac{\Lambda^2}{2}\Box\right)^2}\;\partial^\nu\left(\partial_\mu
J^\mu(x)\right)\right].
\end{equation}
In the next section, we will study the free space solutions of the
inhomogeneous infinite derivative Proca equation.

\section{Free Space Solutions of the Infinite Derivative Abelian Proca Model }
In free space $(i.e.,\; J^\mu =(0,\underbrace{0,\ldots,0}_{D\;\;times}))$, the
infinite derivative field equation (29) can be written as follows:
\begin{equation}
\frac {1}{\left(1-\frac{\Lambda^2}{2}\Box\right)^2}\Box A^\nu (x)+
\left(\frac{mc}{\hbar}\right)^2A^\nu(x)=0.
\end{equation}
The modified field equation (30) has the following plane wave
solution
\begin{equation}
A^\nu (x)=A\;e^{-\;\frac{i}{\hbar}\textbf{p}\cdot
\textbf{x}}\;\epsilon^\nu(p),
\end{equation}
where $\epsilon^\nu(p)$ is the polarization vector and $A$ is the
amplitude of the vector field. If we insert (31) in (30), we will
obtain the following modified dispersion relation:
\begin{equation}
\frac{\textbf{p}^2}{\left(1+\frac{\Lambda^2}{2\hbar^2}\textbf{p}^2\right)^2}=m^2c^2.
\end{equation}
The modified dispersion relation (32) leads to the following
modified energy-momentum relations:
\begin{eqnarray}
E^2_{_+}(\Lambda)&=&{\cal M}^2_{_+}(\Lambda)c^4+c^2\vec{\textbf{p}}^2,\\
E^2_{_-}(\Lambda)&=&{\cal M}^2_{_-}(\Lambda)c^4+c^2\vec{\textbf{p}}^2,
\end{eqnarray}
where the effective masses ${\cal M}_{_{+}}(\Lambda)$ and ${\cal
M}_{_-}(\Lambda)$ are defined as follows:

\begin{eqnarray}
{\cal
M}_{_+}(\Lambda)&:=&\frac{2m}{1-\sqrt{1-2\left(\frac{mc\Lambda}{\hbar}\right)^2}},\\
{\cal
M}_{_-}(\Lambda)&:=&\frac{2m}{1+\sqrt{1-2\left(\frac{mc\Lambda}{\hbar}\right)^2}}.
\end{eqnarray}
In order to avoid imaginary masses in (35) and (36) the
characteristic length scale $\Lambda$ must satisfy the following
relation
\begin{equation}
\Lambda \le \frac{1}{\sqrt{2}}\;\lambdabar_{_C},
\end{equation}
where $\lambdabar_{_C}=\frac{\hbar}{mc}$ is the reduced Compton
wavelength of the particle $m$.\footnote{For
$\Lambda=\frac{1}{\sqrt{2}}\;\lambdabar_{_C}$ both effective masses
${\cal M}_{_+}(\Lambda)$ and ${\cal M}_{_-}(\Lambda)$ have the same
value ${\cal M}_{_+}(\Lambda)={\cal M}_{_-}(\Lambda)=2m$.} \\
According to Eq. (37), the maximum value of the characteristic
length $\Lambda$ is
\begin{equation}
\Lambda_{_{max}}=\frac{1}{\sqrt{2}}\lambdabar_{_C}.
\end{equation}
Now, let us study the low-energy (large-distance) behavior of the
effective masses ${\cal M}_{_\pm} (\Lambda)$ for
$\Lambda\longrightarrow 0$. For $\Lambda\longrightarrow 0$ the
effective  masses ${\cal M}_{_\pm} (\Lambda)$ in Eqs. (35) and (36)
have the following low-energy expansions:
\begin{eqnarray}
{\cal M}_{_+}(\Lambda)&=&\underbrace{2m\lambdabar_C
^2\;\frac{1}{\Lambda^2}}_{infinite\;\;
term}\underbrace{-m-\frac{m}{2\lambdabar_C^2}\;\Lambda^2+{\cal
O}(\Lambda^4)}_{regular\;\;terms},\\
{\cal
M}_{_-}(\Lambda)&=&m\left[1+\frac{\Lambda^2}{2\lambdabar_C^2}+\frac{\Lambda^4}
{2\lambdabar_C^4}\;+{\cal O}(\Lambda^6)\right].
\end{eqnarray}
Equations (39) and (40) show that the low-energy limit of our model
contains two massive vector particles, one with the usual mass $m$
and the other a heavy-mass particle of mass
$\frac{2m\lambdabar_C^2}{\Lambda^2}$.\footnote{ It is necessary to note that
the appearance of such heavy-mass particles in higher-order
derivative quantum field theories leads to an indefinite metric (for a review, see
Refs. [23-25]).}

\section{Summary and Conclusions}
In 2017, G. P. de Brito and his collaborators proposed a covariant
generalization of the Kempf-Mangano algebra in a $(D+1)$-dimensional
Minkowski space-time [13]. In this paper, after reformulation of the
Abelian Proca model from the viewpoint of the covariant
generalization of the Kempf-Mangano algebra, we showed that our
modified Proca model describes two massive vector particles (see
Eqs. (35) and (36)). We proved that there is a characteristic length
scale $\Lambda=\hbar l$ in the modified Proca theory (Eq. (19))
whose upper limit is given by Eq. (38), i.e.,
$\Lambda_{_{max}}=\frac{1}{\sqrt{2}}\lambdabar_{_C}$. According to
our calculations, for $\Lambda\rightarrow0$ the modified Proca
theory in Eq. (19) behaves like an Abelian massive Lee-Wick model,
i.e.,
\begin{eqnarray}
\nonumber{\cal L}_{_{Abelian\;\; massive\;\; Lee-Wick\;\;
model}}&=&{\cal L}_0+\Lambda^2{\cal
L}_1+{\cal O}(\Lambda^4)\\\nonumber&=&-\;\frac{1}{4\mu_{0}}F_{\mu
\nu}(x)F^{\mu\nu}(x)-\;\frac{1}{4\mu_ {0}}\Lambda^2 F_{\mu
\nu}(x)\Box F^{\mu
\nu}(x)\\&\,&+\frac{1}{2\mu_0}\left(\frac{mc}{\hbar}\right)^2A_\mu(x)
A^\mu(x)- J^{\mu}(x) A_\mu(x)+{\cal O}(\Lambda^4).
\end{eqnarray}
Now, let us estimate the numerical value of
$\Lambda_{_{max}}=\frac{1}{\sqrt{2}}\lambdabar_{_C}=\frac{1}{\sqrt{2}}\frac{\hbar}{mc}$
 in Eq. (38).
 \\The usual Proca equation (27) plays a fundamental role in nuclear
 and low-energy particle physics [26,27].
The four-dimensional Proca wave equation for a neutral $Z^0$ boson in a nucleus is
\begin{equation}
\left[\frac{1}{c^2}\frac{\partial^2}{\partial
t^2}-\nabla^2+\left(\frac{M_{_Z}c}{\hbar}\right)^2\right]\phi_{_Z}(\vec{\textbf{x}},t)=\frac{1}{\epsilon_0}\rho_{_Z}
(\vec{\textbf{x}},t),
\end{equation}
where $\rho_{_Z}(\vec{\textbf{x}},t)$ is the neutral weak-charge
density (see page 10 in Ref. [26]). The mass of the $Z^0$ boson is
[26]
\begin{equation}
M_{_Z}=(91.187\pm0.007)\quad GeV /
c^2\thicksim100\;proton\;\;masses.
\end{equation}
According to Eq. (38) the maximum value of the characteristic length
scale $\Lambda$ in our paper is proportional to
$\lambdabar_{_C}=\frac{\hbar}{mc}$, i.e.,
\begin{equation}
\Lambda_{_{max}}\sim \;\lambdabar_{_C}.
\end{equation}
Inserting (43) into $\lambdabar_{_C}=\frac{\hbar}{mc}$, we find
\begin{equation}
\lambdabar_{_Z}=\frac{\hbar}{M_{_Z}c}\approx2\times10^{-18}\; m.
\end{equation}
A comparison between Eqs. (44) and (45) shows that the maximum value
of the characteristic length scale $\Lambda$ in this research is
\begin{equation}
\Lambda_{_{max}}\sim 10^{-18}\; m.
\end{equation}
It is interesting to note that the numerical value of
$\Lambda_{_{max}}$ in Eq. (46) is near to the electroweak length scale
[27,28], i.e.,
\begin{equation}
\Lambda_{_{max}}\sim l_{_{electroweak}}\sim10^{-18}\; m.
\end{equation}
On the other hand, the numerical estimation of $\Lambda_{_{max}}$ in
Eq. (46) is about three  orders of magnitude smaller than the
nuclear scale of $10^{-15}\;m$ [28], i.e.,
\begin{equation}
\Lambda_{_{max}}\sim 10^{-3}\;l_{_{nuclear\;scale}}.
\end{equation}
We showed that in the infrared region, the usual Proca model is
recovered (Eq.(15)), while in the ultraviolet region the Abelian Proca model must be
described by an infinite derivative Lagrangian density (Eq. (19)). In our future works, we will study the interparticle potential energy for the modified Proca model which has been presented in this paper.




\end{document}